\documentclass[aps,pra,preprint,superscriptaddress,floatfix,showpacs,showkeys]{revtex4}
\usepackage{graphicx}
\usepackage{rotating}
\usepackage{float}
\usepackage{lineno}
\usepackage{mathtools}

\newcommand{\comment}[1]{}

\begin{document}



\title{Signatures of Discrete Breathers in Coherent State Quantum Dynamics}


\author{Kirill Igumenshchev}
\email[]{kigumens@mail.rochester.edu}
\affiliation{Department of Chemistry, University of Rochester, Rochester, NY  14627}%
\author{Misha Ovchinnikov}
\email[]{ovchinnikov@chem.rochester.edu}
\affiliation{Department of Chemistry, University of Rochester, Rochester, NY  14627}%
\author{Panagiotis Maniadis}
\email[]{maniadis@lanl.gov}
\affiliation{Theoretical Division, Los Alamos National Laboratory, Los Alamos, NM 87545}%
\author{Oleg Prezhdo}
\email[]{oleg.prezhdo@rochester.edu}
\affiliation{Department of Chemistry, University of Rochester, Rochester, NY  14627}%


\date{\today}

\begin{abstract}
Discrete breathers (DBs) -- a spatial time-periodic localization of energy -- are predicted in a large variety of non-linear systems.  Motivated by the conceptual bridging of the DBs phenomena in classical and quantum mechanical representation, we study their signatures in the dynamics of a quantum equivalent of a classical mechanical point in a phase space -- a coherent state.  We show that in contrast to a point in a phase space that can exhibit either delocalized or localized motion, a coherent state can show signatures of a quantum equivalent of both localized and delocalized behavior.  In classical mechanics, the separation between the energy regions of localized and delocalized motion is a point.  In quantum mechanics, this point becomes a \emph{transient region}, in which both tunneling and non-tunneling modes are present.  Furthermore, the \emph{transient region} contains modes that cannot be characterized as either, because the transition from non-tunneling to tunneling modes is smooth.  With a further analysis we document four intriguing observations: 1.  Considered as a function of coupling, the eigenstates go through avoided crossings between tunneling and non-tunneling modes.  2.  The dominance of tunneling modes in high non-linearity region is compromised by an appearance of new types of modes --  \emph{high order tunneling modes}.  These modes are similar to the tunneling modes but have attributes of non-tunneling modes.  3.  There are types of excitations, which prioritize higher order tunneling modes and allows us to study their properties.  4.  For high non-linearity, the auto-correlation function decreases fast; therefore, short-times dynamics is sufficient for modeling quantum DBs.  This work provides a foundation for implementation of modern semi-classical methods to model quantum DBs, bridging concepts of classical and quantum mechanical signatures of DBs, and understanding spectroscopic experiments that involve a coherent state.

\end{abstract}


\maketitle



\section{Introduction}
\label{chap:intro}


Understanding the mechanism of quantum discrete breathers (QDBs) is
important for a number of application areas, including medicine
~\cite{PhysRevE.83.011904},
nano-materials~\cite{PhysRevA.75.063808,PhysRevLett.90.044102,PhysRevB.79.134304}
and quantum computers~\cite{PhysRevLett.84.745}.  From the classical
mechanics perspective, discrete breathers (DBs) is a fairly well
understood phenomena, and occur when energy localizes in the system for
an infinite amount of time. The effect is due to an interplay of
non-linearity of the sites' potentials and coupling between the sites.
In quantum mechanical perspective, the phenomena is different;
a localization in quantum symmetric potential is not possible due to
tunneling.  Experimentally, one can observe traces of QDBs breathers in
the form of anomaly long localization
time~\cite{1751-8121-43-18-183001}.  QDBs have been observed in
crystals~\cite{PhysRevLett.90.044102,PhysRevLett.82.3288,PhysRevB.79.134304},
Bose-Einstein
condensates~\cite{\comment{PhysRevA.73.023607,PhysRevA.16.1692,
    PhysRevA.82.013606,PhysRevA.66.033610,}PhysRevLett.92.230401} and
interacting Josephson
junctions~\cite{PhysRevLett.84.745\comment{,PhysRevB.37.9325}}.
Complimenting the experiments, theoretical solutions of QDBs stems from
quantizing classical solutions for both integrable
\cite{aubry1997breathers} and non-integrable systems
\cite{Bohigas199343}.  Theoretically, DBs have been studied in a number
of systems of a practical importance: molecules~\cite{PhysRevA.43.5518},
biological polymers~\cite{PhysRevE.83.011904}, quantum
dots~\cite{PhysRevA.75.063808}, nano-tubes~\cite{PhysRevB.77.024307} and
graphene sheets~\cite{PhysRevB.81.165418}.


Investigating the quantum dynamics of a coherent state targets two areas
of research: bridging classical and quantum description of DBs and
building a foundation for numerical methods that use coherent states.
A coherent state is a quantum equivalent of a point in a classical phase
space.  Comparing signatures of DBs in dynamics of a point in
a classical phase space and a coherent states, outlines the conceptual
differences and similarities between classical and quantum descriptions
of DBs.  Even numerically, large experimentally accessible systems are
difficult to model using exact quantum dynamics methods.  One needs
approximations that keep track of quantum effects and scale well with
system size.  A number of  methods that rely on time-evolution of
a coherent state, such as
QHD~\cite{prezhdo:4450\comment{,pereverzev:134107,heatwole:244111,prezhdo:2995,springerlink:10.1007/s00214-005-0032-x,doi:10.1021/jp0709050,Miyachi2006585,Horsfield200816,Ando2003532,heatwole:244111,pahl:8704,prezhdo:2995}},
Herman-Kluk (HK) propagator~\cite{Herman198427} and its higher order
extensions~\cite{hochman:061104,Kay20063}, coupled coherent states (CCS)
~\cite{\comment{halashilin:5367,}shalashilin:244111},
matching-pursuit/split-operator Fourier transform
(MP-SOFT)~\cite{wu:1676,wu:6720} and coherent-state path-integral
(CSPI)~\cite{PhysRevE.69.066204}, are effective for large systems. These
methods have been successfully used only for slightly non-linear
systems; in addition, these methods were derived to be applied to
harmonic or slightly anharmonic system. Modeling highly non-linear
phenomena such as DBs can be a challenge. With an exception of
QHD~\cite{igumqhd}, these methods have not been applied to QDBs.
Studying the quantum dynamics of a coherent state will show the
limitations of these techniques when applied to QDBs. 


The current study is related to previous work on modeling DBs in systems
that may have quantum effects, fundamental theory on QDBs, and the
applicability of semi-classical methods.  Due to the high computational
cost of quantum simulation, systems of practical importance that contain
quantum
effects~\cite{PhysRevA.43.5518,PhysRevE.83.011904,PhysRevA.75.063808,PhysRevB.77.024307,PhysRevB.81.165418}
are simulated with classical dynamics. Our work illustrates limitations
of classical dynamics by outlining quantum effects in the dynamics of
a coherent state.  Previous theoretical work on QDBs manifests from the
quantization of an integrable~\cite{aubry1997breathers} and
non-integrable~\cite{Bohigas199343} dimer, the analysis of a splitting of
QDBs modes~\cite{PhysRevE.58.339} and existence of an avoided
crossing~\cite{0953-8984-9-33-007}.  We build on this work by providing
an analysis of coherent states rather than eigenstates.  Conventionally
QDBs are studied by finding eigenstates of the system's
Hamiltonian~\cite{PhysRevLett.76.1607}. Since here we are interested in
the dynamics of a coherent state rather than eigenstates, our approach
involves simulating the dynamics of a superposition of eigenstates in a form
of a coherent state. Motivation for this project came from our work on
applying QHD~\cite{igumqhd} and HK~\cite{igumhk} semi-classical
methods to model QDBs; in these two works, the influence of tunneling
modes (a quantum mechanical equivalent of DB modes) on the quantum
dynamics of coherent state was unresolved. 


Here, we focus on the fundamental concepts of QDBs in coherent state
dynamics that build foundation to practical applications.  Our approach
is to study the dynamics of a coherent state in the simplest quantum
mechanical Hamiltonian that accounts for QDBs phenomena.  We choose
a two-site quartic potential with a linear coupling. To study the
transfer and localization of energy, we start by displacing a coherent
state on one of the sites, while keeping the other site in a ground
state.  Analysis of the spectrum will provide insight into the behavior
of the coherent state. Contour plots of the distribution of the
eigenstate wave-function between the sites will give a more detailed
analysis of energy between the sites for a specific eigenstate. Spectrum
and contour plots help us to draw conclusions regarding the earlier
mentioned problems.

The paper is structured in following way: In the following section we
discuss methods and numerical procedures. The results section starts
with an analysis of the dynamics of the coherent state by investigating
the spectrum. The second part of the results shows a transient region,
a region with both tunneling and non-tunneling modes. The last part of
the results briefly discusses existence of the following effects:
avoided crossing between tunneling and non-tunneling modes,
manifestation of higher order tunneling modes (HOTM), control of HOTM
intensity with initial displacement, and rapid decrease of
autocorrelation function due to non-linearity.  The last section will
outline our findings and their importance. 

\section{System and Methods}

Breathers appear in a wide range of systems. We consider a simple model
system that is computationally accessible, conceptually simple and
congruent with natural systems: a Hamiltonian with quartic oscillators
and linear coupling: 
\begin{eqnarray} H = c_h (P_1^2 + P_2^2 + X_1^2 + X_2^2) + c_a (X_1^4
  + X_2^4) +c_c X_1 X_2 \label{eq:ham} \end{eqnarray}
Here, $P_1$, $P_2$, $X_1$ and $X_2$ are either classical coordinates or
quantum mechanical operators, $c_h$ is a harmonic constant with the value
"$\frac{1}{2}$", $c_a$ is responsible for the anharmonicity of the
potential, and $c_c$ is a linear coupling coefficient. 

The phenomena of DBs and QDBs are due to non-linearities in the
potential.  When the oscillators are in low energy states, the quadratic
term dominates over the quartic term -- the motion is harmonic. There
is an exchange of energy between the oscillators. However, suppose one
of the oscillators is higher in energy. The quartic term dominates over
the quadratic term. In both classical and quantum mechanical
perspectives, the resonant condition between the sites disappears. In
classical mechanics, the energy does not transfer onto the other
oscillator.  In quantum mechanics, there is energy transfer due to
tunneling but only to the same energy levels on the other oscillator.

For our calculations, the Hamiltonian is best represented in the
occupational basis set of harmonic oscillators, also known as Fock
states.  The occupational basis set is a direct product of the
occupations of all oscillators: $ |N_1 N_2 \cdots N_i \rangle$, where
$ N_i$ is the occupational state on the $ i^{th}$ oscillator.  For our
calculations, $i=2$. The Hamiltonian of the system in terms of creation
and annihilation operators of the 1D harmonic oscillators is given as:
\begin{eqnarray} H&=& c_{h} * H_{h} + c_{a} * H_{a} + c_{c} * H_{c}\\
  H_{h} &=& \sum_i a_i^\dagger a_i + a_i a_i^\dagger \\ H_{a} &=& \sum_i
  a_i a_i a_i a_i + a_i a_i a_i a_i^\dagger + a_i a_i a_i^\dagger a_i
  + \cdots + a_i^\dagger a_i^\dagger a_i^\dagger a_i^\dagger \\ H_{c}
  &=& \sum_i \left( a_i a_{i+1} + a_i a^\dagger_{i+1} + a^\dagger_i
    a_{i+1} + a^\dagger_i a^\dagger_{i+1} +\cdots + a^\dagger_{i+1}
    a^\dagger_i \right) \label{eq:hamSecondQuant} \end{eqnarray}
$c_{h} = 1.0$, $c_a$ is a coefficient for anharmonicity in the system
(for example $c_a=0.02$ in most of our calculations) and $ c_{c}$ is
a coefficient for coupling between the oscillators.

The number of levels in the basis set and the size of the matrix
required for the calculation depend on the displacement of the coherent
state and anharmonicity of the system.  At the higher energy, where the overlap
between initial state and eigenstates is small, the basis set is
truncated.  For example, truncating as early as 10 occupational states
on each site is enough for the displacement of 1.0 a.u. and $c_a<0.05$.
The total dimensionality is the number of states of a single oscillator
to the power of the number of the oscillators.

We choose a coherent state as an initial state because it helps to
bridge the gap between understanding physics in the classical and
quantum limits and it helps to bring the theory closer to the experiment
by using coherent state base methods. A coherent state is a state of a minimum uncertainty~\cite{landau3}.  In
the basis set of occupational eigenstates of the harmonic oscillator $|i
\rangle$, a coherent state of a single site takes the form of
Eq.\ref{eq:cohState}~\cite{landau3}.
\begin{eqnarray}
  |\alpha (x_0)  \rangle &=& \sum_i c_i (x_0) |i \rangle, \\
  c_i(x_0) &=& e^{-\frac{|x_0|^2}{2}} \frac{|x_0|^i}{i!}, \nonumber
  \label{eq:cohState}
\end{eqnarray}
where $c_i(x_0)$ represents the contribution from the $i^{th}$ energy
level and $x_0$ is the displacement of the coherent state. Our basis set
is a direct product of occupational states of two harmonic oscillators. 
\begin{eqnarray}
  |\psi (\mathbf{x_0})\rangle &=& \sum_{i,j} c_i (x_{0i}) c_j  (x_{0j})|i j\rangle \\
  c_i (x_{0i}) &=& e^{-\frac{|x_{0i}|^2}{2}} \frac{|x_{0i}|^i}{i!} \nonumber\\
  c_j (x_{0j}) &=& e^{-\frac{|x_{0j}|^2}{2}} \frac{|x_{0j}|^j}{j!} \nonumber 
  \label{eq:multicohState}
\end{eqnarray}
To resemble the transfer of energy in classical dynamics, the coherent state
is displaced on only one of the oscillators. Now that we have defined
the initial state and know the eigenstates of the system, we can observe
the dynamics.

The spectrum represents the dynamics of the system in the frequency
domain.  The spectrum of a quantum system consists of discrete energy
values. These energy values are found by diagonalization of the
Hamiltonian. For diagonalization, we use the symmetric QR routine from
GMM++ library (an interface to LAPACK), which produces the eigenvalues
and eigenvectors of the system. The intensity of the spectral lines
depends on the initial state. One can either do a Fourier transform of
the auto-correlation function or write an initial state in the
eigenstate basis. We use the second approach. We find the intensity of
the eigenvalue's peak by taking a normalized dot product between the
initial state and the corresponding eigenstate (Eq.\ref{eq:spectrInt}).
\begin{eqnarray} c_k^{initial}=c_k^{final} = \frac{\langle
    \psi_{initial} | e_k \rangle }{\langle \psi_{initial}
    | \psi_{initial} \rangle}, \label{eq:spectrInt} \end{eqnarray} where
$|e_k\rangle$ is the $i^{th}$ eigenstate. 

To view the dynamics in the time domain, we compute the auto-correlation
function, which is a projection of the initial state onto the final one,
($\langle \psi_{final} | \psi_{initial} \rangle$). To get the final
state from the initial one, we need to use a time-propagation operator
$e^{-i\hat{H}t}$. To do that, we expand the initial state in terms of
eigenstates.  Eq. \ref{eq:corr} expresses the auto-correlation function in
terms of the eigenstates and coefficients $c_k$, which are a projection
of the initial state onto the eigenstates (Eq.
\ref{eq:spectrInt}).
\begin{eqnarray}
  \langle \psi_t | \psi_0 \rangle = \sum_k |c_k|^2 e^{-i E_k t}
  \label{eq:corr}
\end{eqnarray}

\section{Results}

\subsection{Tunneling and non-tunneling modes}

\begin{figure}[H]
  \begin{center}
    \includegraphics{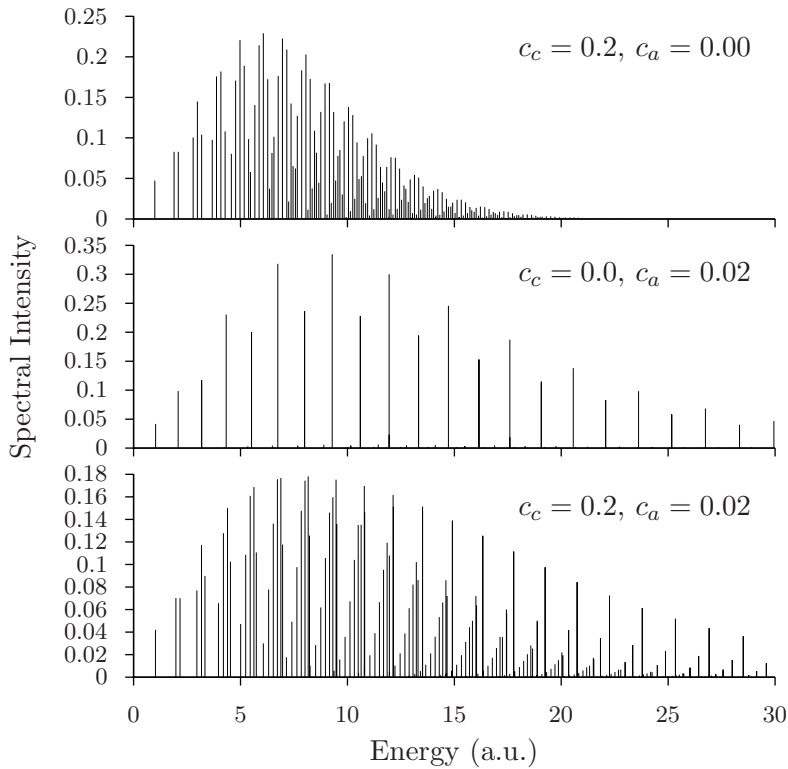}
  \end{center}
  \caption{Spectrum of a coherent state in a system with the Hamiltonian
    in Eq.\ref{eq:ham}. The {\bf top} panel is the harmonic limit
    ($c_a=0$).  All of the spectral lines are non-tunneling modes. The spectral
    lines in the {\bf middle} panel are tunneling modes --- the anharmonic
    limit ($c_c=0$). The {\bf bottom} panel shows modes in a coupled
    anharmonic potential. The spectrum
    (bottom) has modes that resemble both limiting cases (top and
    middle). The modes that resemble the delocalized modes are
    \textit{non-tunneling modes} and the modes that resemble the localized
    modes are \textit{tunneling modes}.}
  \label{FIGcouunh}
\end{figure}

Signatures of breathers can be observed in the spectrum of the system.
In the spectrum, we would like to distinguish two types of modes:
\textit{non-tunneling} and \textit{tunneling} modes.  Tunneling modes
arise when the coupling between the oscillators is not strong enough to
create an eigenstate, where the wave function density is shared between
the sites -- the particle has to tunnel from one site to the other. In
a classical limit, the tunneling time approaches infinity and these
tunneling modes become DB modes. When there is a considerable amount of
shared wave-function density between the sites, an alternative type of
modes manifests -- non-tunneling modes.  These modes can be seen clearly
in a low energy region, where coupling dominates. Besides limiting
cases, where there is either a small or large amount of shred density
between the sites, it is sometimes hard to distinguish between tunneling
and non-tunneling modes. In the following sections we show that the
transition between these two types of modes is smooth. To have a better
understanding of how the two coexist in the same energy region, let us
first focus on the limiting cases: coupled harmonic oscillators with no
anharmonicity (Fig.~\ref{FIGcouunh} (top)) and anharmonic oscillators
with no coupling (Fig.~\ref{FIGcouunh} (middle)). We then analyze the
case where the potential combines both anharmonicity and coupling
(Fig.~\ref{FIGcouunh} (bottom)).

In the coupled case (Fig.~\ref{FIGcouunh} (top)), we see a splitting of
the energy levels. All of the eigenstates in this case are
non-tunneling. For example, at 2 a.u., a system of uncoupled harmonic
oscillator would have the second lowest eigenstate $|1 \rangle$. Since there is
a coupling between the oscillators, these $|1 \rangle$ states of each
oscillator combine into a linear combination of symmetric and
anti-symmetric states: $|10\rangle + |01\rangle$ and $|10\rangle
- |01\rangle$. For the third lowest states ($|02\rangle$, $2\times
|11\rangle$ and $|02\rangle$), we also see splitting around the energy
of $|2\rangle$ of a system of uncoupled harmonic oscillator at 3 a.u.
For each eigenvalue of a 1D oscillator, in the case of coupled
oscillators, we get a group of states.  Each state in that group is
a linear combination of states that have the same energy; for example,
$|05\rangle$, $|14\rangle$, $|23\rangle$, $|32\rangle$, $|41\rangle$ and
$|50\rangle$. This is an example of what we call \emph{same-quanta} group --
a group of states that have the same energy. We extend this definition
to non-linear cases by defining \emph{same-quanta group} as a group of
states that become degenerate when non-linearities of the potential are
decreased to zero.

The other limiting case is when there is anharmonicity and no coupling
(Fig.~\ref{FIGcouunh} (middle)). The spectrum is similar to that of a 1D
oscillator. There is no splitting between the states of the different
sites. The ground state energy is at 1 a.u.; this is because there are
two oscillators with energy $\approx 0.5$ a.u. for each. Since there is no
coupling, all of the modes on the graph (Fig.~\ref{FIGcouunh} (middle))
are tunneling modes in the limit of tunneling time being infinity.

In a coupled harmonic case, the levels split due to coupling. A similar
situation occurs in the case of coupled anharmonic potential. Once there
is a coupling between tunneling modes, they split.  The difference is
that in the anharmonic case only two states can be involved in
splitting; for example, $|90\rangle$ and $|09\rangle$ are involved and
$|81\rangle$, $|72\rangle$, $|63\rangle$, ..., $|18\rangle$ are not
involved in the same splitting. In the harmonic coupled case, all of the
levels with the same number of quanta split together.

The system that includes both coupling and anharmonicity
(Fig.~\ref{FIGcouunh} (bottom))resembles both limiting cases: $c_a=0$
(Fig.~\ref{FIGcouunh} (top)) and $c_c=0$ (Fig.\ref{FIGcouunh} (middle)).
At lower energy, the non-tunneling modes dominate and at higher energy,
tunneling modes dominate.  A classical mechanical description of DBs
separates localized and delocalized modes' energy regimes by separatrix.
Our analysis of the eigenstates contour plots shows that, from a quantum
mechanical perspective, non-tunneling  modes, the quantum equivalent of
delocalized modes, can be found higher in the energy spectrum.  There is
also no clear separation between regions with non-tunneling and
tunneling modes. As energy increases, tunneling modes slowly emerge out
of non-tunneling modes that are lower in energy.  This region in energy,
where both tunneling and non-tunneling modes can be found, we define as
a \emph{transient region} (Fig.~\ref{FIGcouunh} (bottom), a region of
10-15 a.u.).  The tunneling modes manifest as isolated peaks that
resemble the modes in uncoupled quartic potential.  At the energy higher
than the transient region, they can be seen clearly. The contour plot
analysis shows that the wave function of the tunneling modes is
localized on the sites. These modes are equivalent to localized modes in
a classical mechanical description of DBs.  The non-tunneling modes
resemble the modes in coupled harmonic potential.  They have
a significant wave-function density between the sites and appear lower
in energy where non-linear contributions are not significant. The
non-tunneling modes are an equivalent of delocalized modes in classical
mechanics description of DBs.


We deliberately choose the values for the $c_a$ and $c_c$ coefficients
and the position of the coherent state on site 1 ($q_1$). The intention
is to clearly show a transient region and the interplay between the
non-tunneling and tunneling modes.  Not many natural systems have an
initial state that spans over such a large number of states.  For
example, in C-H, stretch anharmonicity dominates the coupling already in
the second excited state~\cite{wyatt:10732}.  However, our choice of the
system parameters and initial conditions gives a general understanding
of the dynamics.  In heavy oscillators, such as nano-mechanical
cantilever arrays or large molecules with hydrogen bonding, the
displaced state can spans over a large number of states, which would be
similar to our model.

The spectrum shows the overlap of the initial state with eigenstates.
Let us understand the mechanism that develops this spectrum.  The
intensity is a projection of an initial state on eigenstates. The more
similar the distribution of the wave-function between initial state and
eigenstate, the higher the intensity at the corresponding eigenvalue.
For example, at low energy and at the very high energy of the spectrum,
the initial state does not have much density; therefore, we do not see
much intensity. The majority of intense states are at the energy
$H(q_1=3.5,q_2=0.0)$.  If only one of the oscillators has a displaced
state and the other one does not, the initial state contour plot has
non-zero elements only along one of the axes.  Non-tunneling eigenstates
at lower energy have a considerable amount of density that overlaps with
non-zero elements of the coherent state.  At higher energy,
anharmonicity prevails and the tunneling eigenstates have a considerable
overlap with the initial state. The wave-function of tunneling
eigenstates concentrates on the oscillators instead of being shared
between them. These tunneling eigenstates correspond to the highest
energy states from the group of states that came from single level of
uncoupled oscillator.  Since our initial state has dominant population
on the oscillators and not shared between them, the eigenstates that
have density localized on oscillators, have higher intensity.


Contour plots of eigenstates show the energy distribution within an
eigenstate. Quantum dynamics of the coherent state (Fig.~\ref{FIGcouunh}
(bottom)) result from the projection of the initial state onto the
eigenstates. By visualizing and comparing the density distribution of
initial wave function and eigenstates, one can understand how the
spectrum forms.  For example, the wave-function of tunneling eigenstates
has higher density concentrating on the sites; it can be seen as higher
overlap with $|i,j \rangle$ states where $|i-j| \approx i+j$.
Therefore, the initial state, which is localized on sites rather than
delocalized between the sites, will have a higher intensity.

\begin{figure}[H]
  \begin{center}
    \includegraphics[scale=0.7]{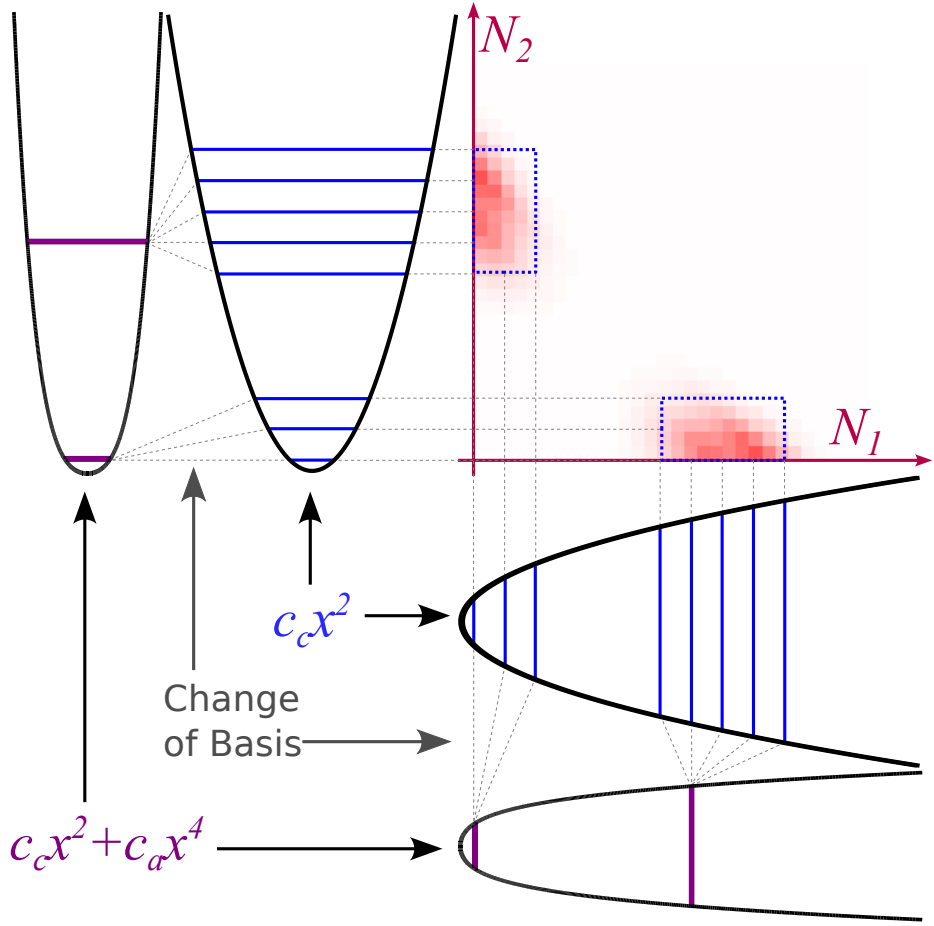}
  \end{center}
  \caption{
    (Color online)
  The schematic shows how to interpret a contour plot of an eigenstate. 
  Each axis represents Fock states for each site's harmonic oscillator
  ($c_c x_i^2$). The value on the axis is the eigenvalue of the harmonic
  oscillator Fock state. The schematics shows that to portray the
  calculated eigenstate, we need to change the basis set from anharmonic
  ($c_c x_i^2 + c_a x_i^4$) to the harmonic ($c_c x_i^2$) one. Due to this
  representation of the eigenstate, we can distinguish the tunneling mode by
  identifying that localization of the wave-function (red) is towards
  the axis.  }
  \label{FIGcontourSchematics}
\end{figure}

2D contour plots of eigenstates' wave function visualize how the
probability density distributes among the oscillators. The procedure for
getting these contour plots is the following. The Hamiltonian matrix is
written in the Fock basis, which is a direct product of occupational
states of each oscillator. The matrix is diagonalized. The eigenstates
that are in the Fock basis are rewritten as a 2D matrix. The columns of
the matrix are enumerated according to the Fock states of the first
oscillators: $|0,i\rangle$,$|1,i\rangle$, \ldots, $|n,i\rangle$; where
$i=0 \cdots n$. The rows are enumerated according to the Fock states
of the second oscillator: $|i,0\rangle$,$|i,1\rangle$, \ldots,
$|i,n\rangle$. And the elements of the matrix are elements of the
eigenstate that is written as a 2D matrix. The contour plots show these
matrices. A value at a coordinate $(x,y)$ corresponds to a component of
$|x,y\rangle$ basis vector. Fig.~\ref{FIGcontourSchematics} gives an
example of such contour plot. The figure also explains how to interpret
the plot in terms of the occupational levels of the harmonic
oscillators. 

These plots help to visualize and distinguish localized from delocalized
states.  There are other benefits, too. The contour plots of the
eigenstates are well suited to illustrate the transition between
localized and delocalized modes and identify modes that have signatures
of both.  We will also describe the origin of the higher order
breathers, the small satellite peaks that are a little lower in energy
then localized modes on Fig.~\ref{FIGcouunh} (bottom).

\begin{figure}[H]
  \begin{center}
    \includegraphics[width=\textwidth]{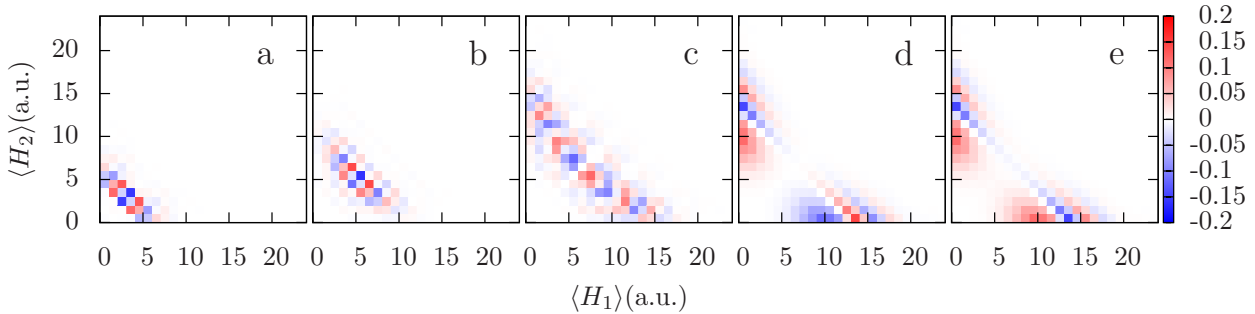}
  \end{center}
  \caption{(Color online)
    Plots \textbf{a}, \textbf{b} and \textbf{c} are examples of
  non-tunneling modes.  They correspond to 21st (with the energy 7.1494
  a.u.), 50th (11.6861 a.u.) and 95th (16.3442 a.u.) energy levels.
  \textbf{d} and \textbf{e} plots are examples of tunneling modes.  They
  are quantum equivalent of classical breathers. The two shown plots are
  anti-symmetric and symmetric states corresponding to 93rd (16.3329
  a.u.) and 94th (16.3332 a.u.) energy levels respectively.}
  \label{FIGnormalLocalExample}
\end{figure}

Fig.~\ref{FIGnormalLocalExample} shows examples of tunneling (d, e) and
non-tunneling modes (a,b,c). Tunneling mode contour plots show
a non-zero density next to the axis, where the value on one axis is high
and on the other axis is low. This density distribution shows that one
of the oscillator is in the lower modes and the other one is in the higher
ones. Another way to distinguish the tunneling modes is to see that the
value of the plotted eigenstate along the rising diagonal is close to
zero.  Non-tunneling modes, on the other hand, have values of
wave-function concentration on the increasing diagonals, meaning the
eigenstate wave-function is shared between both sites.

The wave-function value between non-tunneling and tunneling eigenstates
oscillates differently along the decreasing diagonal -- the line
connecting $|0,n\rangle$ and $|n,0\rangle$. In tunneling modes, the wave
function either does not change sign at all along the decreasing
diagonal for the symmetric case, or changes the sign only once for the
anti-symmetric case. In the tunneling modes, the wave function varies
slightly. In non-tunneling modes on the other hand, the wave function
quickly varies, which is seen as blue and yellow peaks alternating along
the diagonal.

In more detail, the contour plot of the tunneling modes
(Fig.~\ref{FIGnormalLocalExample} (d,e)) correspond to the pair of peaks
at 16.33 found in the spectrum plot (Fig.~\ref{FIGcouunh} (bottom)). The
pair appears as a single peak; two states are only 0.003 a.u. apart. The
eigenstate that is higher in energy is symmetric and the lower one is
anti-symmetric. The small splitting between symmetric and anti-symmetric
states of the tunneling modes pair is consistent with classical
formulation of DBs.

Tunneling modes become breather modes in the classical limit ($xp\gg
\hbar$).  An important observation is that the distribution of absolute
wave-function is very similar in the pair but one state is symmetric and
the other one is anti-symmetric. From these two observations an
important similarity to the classical mechanical view can be concluded.
If both states are occupied, then the density on one of the oscillators
cancels out.  However, because these have different energies, they
evolve in time with different periods and at some point there will be
more energy on the other oscillator. This contradicts the classical
behavior of breathers.  However, the splitting is very small and
therefore the time of the transfer is very long.

Illustrating the intensity and distribution of energy we understand how
the energy is distributed between the sites. This allows us to study in
detail the transient region, the variation of coupling, the influence of
initial displacement and the higher order tunneling modes.  Contour
plots proved to be an important tool to distinguish between localized
and delocalized modes. They would be even more important when the
density of states increases and the spectrum becomes complex; for
example, in soft potentials or when $xp\gg \hbar$. Furthermore, contour
plots help us to understand the evolution of non-tunneling into
tunneling modes.

\subsection{Transient Region}
\label{sec:transition}


Most of the potentials in nature are non-linear and have localized modes
when the energy is high enough. At lower energies the non-linear
potential can be approximated by linear ones. This approximation is
similar to the sine approximation for the motion of a pendulum that one
learns in an introductory physics classes.  With an increase in energy,
the non-linearity becomes more important and the system dynamics changes
from delocalized modes to localized modes.  The transition is important
because the behavior of the system changes significantly. In the
delocalized mode regime, the system transfers the energy from one site
onto the next one but in the localized modes regime the energy is
localized in the classical limit and takes almost infinite time to
transfer in the quantum limit\comment{footnote{In quantum limit,
a symmetric system has symmetric eigenstates and a non-symmetrical
localization is not possible. However if there is initial localization,
the transfer of energy will happen through tunneling.  The tunneling
time increases fast as anharmonicity dominates over coupling.}}. We
model this transition in the simplest potential -- two coupled quartic
oscillators. In other potentials the transition might be quantitatively
different but conceptually  the same.


The transition from the non-tunneling modes region to the tunneling
modes region has a different mechanism in classical and quantum
mechanics. In classical mechanics the two regions are divided by
a separatrix. In contrast, quantum mechanics cannot have such a sharp
division. The energy, in quantum mechanics, does not localize -- it
takes an almost infinite amount of time to transfer through tunneling.
Therefore, in contrast to classical mechanics, tunneling modes slowly
evolve from the non-tunneling modes as energy increases. Understanding
the quantum equivalent for the transition from delocalized to localized
modes will help to understand the limitations of a classical model,
while providing insight into semi-classical dynamics, and complimenting
experiments on QDBs.


\begin{figure}[H]
  \begin{center}
    \includegraphics[width=\textwidth]{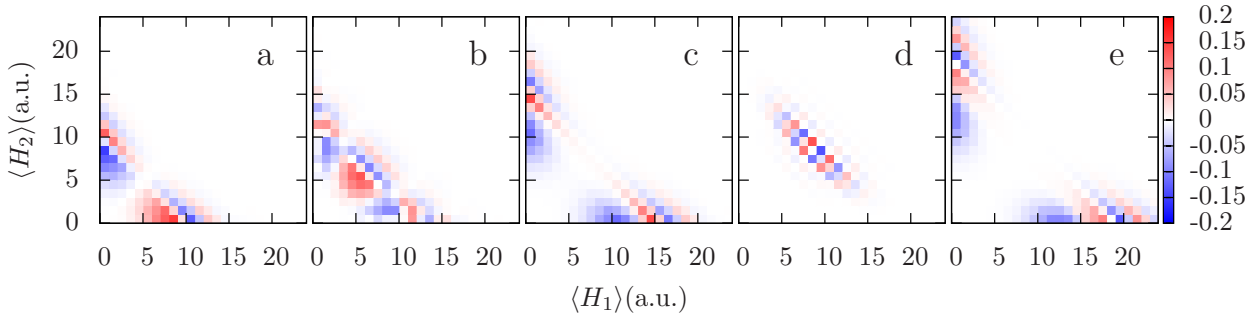}
  \end{center}
  \caption{ (Color online) 
  Contour plots (see Fig.~\ref{FIGcontourSchematics}) of delocalized (\textbf{b}, \textbf{d}) and tunneling 
  (\textbf{a}, \textbf{c},  \textbf{e}) modes within the transient region
  (12-23 a.u.). The plots are in the order of increasing energy: 
  plot \textbf{a} is 55th state with the energy 12.1386, \textbf{b} --
  75th -- 14.454, \textbf{c} -- 110th --17.7796, \textbf{d} --145th --
  20.82 and \textbf{e} -- 163 -- 22.2542. The figure shows that in the
  transient region, the tunneling and the non-tunneling modes alternate.  }
  \label{FIGevalt}
\end{figure}


Depending on the system's Hamiltonian and the initial conditions,
transient region can contain part or all of the coherent state.  There,
non-tunneling and tunneling modes co-exist. Fig.~\ref{FIGcouunh}
(bottom) shows that they alternate with an increase in energy.
Tunneling modes are the highest energy modes from the same-quanta group.
The transient region may contain a number of same-quanta group.  It can
also contain non-tunneling states followed by the tunneling states of
the same-quanta group, further followed by non-tunneling states and
tunneling states of the next same-quanta group.  Fig.~\ref{FIGevalt}
gives an example of alternating tunneling and non-tunneling modes.

\begin{figure}[H]
  \begin{center}
    \includegraphics[width=\textwidth]{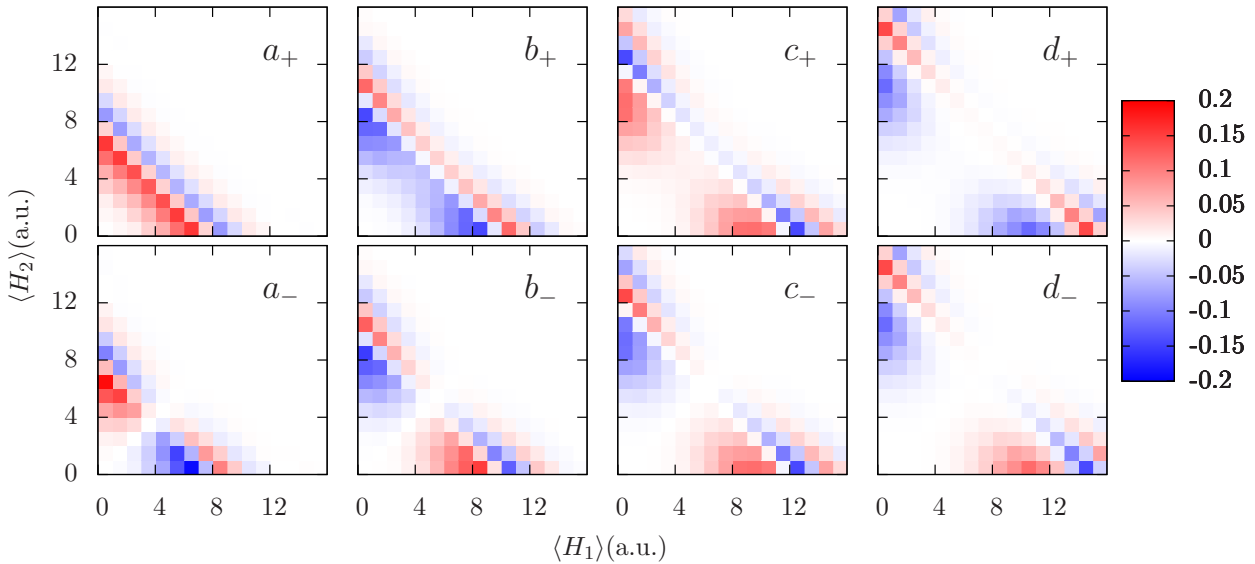}
  \end{center}
  \caption{(Color online) 
  Smooth transition from non-tunneling modes (\textbf{a}) into tunneling
  modes (\textbf{d}). Vertical pairs of plots show symmetric
  ($\textbf{+}$) and anti-symmetric ($\textbf{-}$) eigenstates. From the
  same-quanta group, these states have the highest localization.  As the
  energy of the states increases (\textbf{a} $\to$ \textbf{d}), the
  wave-function moves towards the axis.  The presence of the
  wave-function between the axis, which corresponds to the density
  between the sites, gradually decreases; \textbf{d} has negligible
  amount of wave-function between the sites -- the wave function hast to
  tunnel from one site to the other. The corresponding values of energy
  (E) and intensity (I) for the plots are:
  $a_+)$ E=9.46539    I=0.175297,
  $a_-)$ E=9.50399    I=0.135935,
  $b_+)$ E=12.1386    I=0.161693,
  $b_-)$ E=12.1478    I=0.151122,
  $c_+)$ E=14.9101    I=0.139205,
  $c_-)$ E=14.9113    I=0.138124,
  $d_+)$ E=17.7795    I=0.11148 and
  $d_-)$ E=17.7796    I=0.111422.
  }
  \label{FIGev_h}
\end{figure}

In classical mechanics, the transition from localized to delocalized
modes is a step-function. Fig.~\ref{FIGev_h} shows highest energy
eigenstates from the same-quanta groups within the transient region.
Here one can observe an iterative transformation from non-tunneling to
tunneling modes. This quantum in nature phenomenon significantly
contrasts with the step transition from delocalized to localized modes
in classical mechanics.

Unlike in a classical mechanical perspective, the transition to
tunneling modes from non-tunneling  modes is smooth and there is no
distinct separatrix as in the classical case.  Fig.~\ref{FIGev_h} shows
this transition. The last two plots on the right ($d_-$ and $d_+$) are
clearly tunneling modes; there is no shared density between the sites
and the wave-function has to tunnel through. The majority of
contribution to these eigenstates comes from basis states that have
excitation only on one of the sites; for example, $|14,0\rangle$ and
$|0,14\rangle$. The two plots on the left of Fig.~\ref{FIGev_h} ($a_-$
and $a_+$) are a pair of eigenstates that resemble the tunneling mode.
The pair of states, $b_-$ and $b_+$ (Fig.~\ref{FIGev_h}) has features
similar to the tunneling modes, it is more delocalized than the
tunneling mode since there is a significant amount of density on the
increasing diagonal, meaning that there is overlap with basis states
that have density delocalized between the sites. The pair in the middle
is between two cases. We can see the development of a localized state.

\subsection{Additional observations}

\subsubsection{Variation of the Coupling}
\label{ssec:scale}

Slight changes in the setup of an experiment can vary the shape of the
potential. For example,  one can change parts of the molecules to be
heavier or lighter, or use a different kind of a solution. These kind
of slight changes can control the interplay between the non-linearity
and coupling in the potential.


From Fig.~\ref{FIGcouunh}, we are familiar with signatures of breathers
in a spectrum.  We extend those results by investigating not only how
they appear but also how they depend on the parameters of the potential.
We analyze the evolution of the eigenvalues with respect to coupling
coefficients in the Hamiltonian (Eq.\ref{eq:ham}).  We start from zero
coupling ($c_c=0$) and increase it to the point where coupling is much
larger than anharmonicity, where the spectrum would resemble the
harmonic coupled spectrum.  We focus on eigenvalues of the same-quanta
group of states and print the eigenvalues between 9.1 and 11.4. The
coefficient of anharmonicity is constant: $c_a=0.02$.

\begin{figure}[H]
  \begin{center}
    \includegraphics[width=\textwidth]{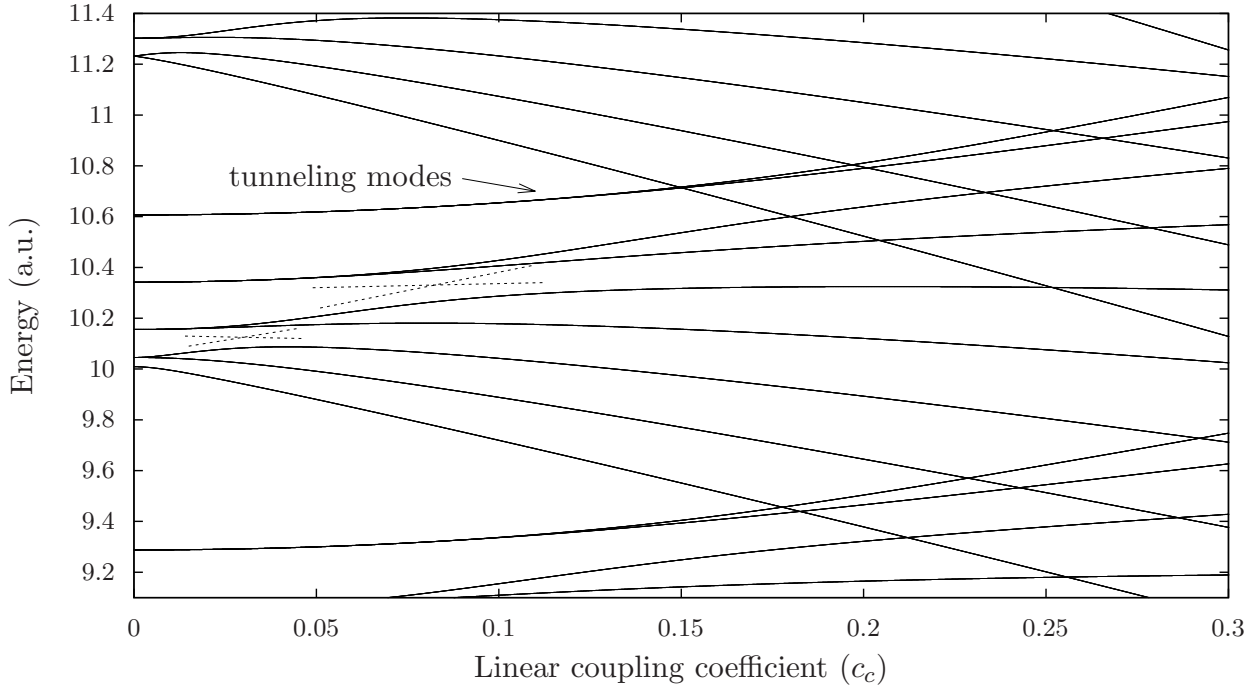}
  \end{center}
  \caption{Eigen energy of quartic dimer (Eq.~\ref{eq:ham}) with
  $c_h=0.5$ and $c_a=0.02$. Dashed crossing lines indicate examples of
  the avoided crossings. 
  }
  \label{FIGcou_unharm_crossing__scale_energy}
\end{figure}

Fig.~\ref{FIGcou_unharm_crossing__scale_energy} shows dependence of
eigenvalues of anharmonic oscillators on coupling ($c_c$). The plot
shows variation of eigenvalues from one extreme of two uncoupled
anharmonic oscillators ($c_c=0.0$) to another that has coupling
($c_c=0.3$) much larger than anharmonicity ($c_a=0.02$).  With this
plot, we would like to focus on a group of same-quanta states.  At
$c_c=0$, the eigenstates of that group have energy in the interval
$\approx (10,10.6)$ a.u. The figure clearly shows how increase in
coupling causes degenerate states to split. The tunneling states, which are the
top states of the same-quanta group, are the last ones to split. Keeping
in mind that the size of splitting is inversely proportional to the transfer
time between the sites and noting that tunneling state approaches a similar
size of splitting as lower-lying non-tunneling modes, one can conclude
that in a quantum mechanical perspective the transition from tunneling to
non-tunneling states is smooth. This observation compliments our result
in section~\ref{sec:transition} on the smooth transition between
tunneling and non-tunneling modes.

We observe avoided crossing for the states from the same-quanta group.
At $c_c\approx 0.8$, the top (symmetric) state that starts at $\approx
10.34$ avoids crossing with the low state that starts at $\approx
10.14$.  Another avoided crossing happened at $c_c \approx 0.04$ for the
next lower pair of states. In harmonic case, there is no crossing since
the degenerate states split linearly and do not cross. Previously,
avoided crossing was observed in anharmonic
trimer~\cite{0953-8984-9-33-007}. The avoided crossing is expected since
the eigenstates belong to the same symmetry group of irreducible
representations. The practical importance of this result would show up
in systems with many degrees of freedom.  In multidimensional systems,
the avoided crossings become conical intersections that allow for
radiation-less transition between the energy
surfaces~\cite{PhysRevA.47.2601}. 

\subsubsection{Higher Order tunneling modes}


In the spectrum plots (Fig.~\ref{FIGcouunh}), one can notice that
besides tunneling modes, there are eigenstates of considerable intensity
in the anharmonic regime, where there are no non-tunneling modes -- the
energy region beyond 16 a.u. (Fig.~\ref{FIGcouunh} (bottom)).These
eigenstates do not classify as non-tunneling modes.  In this higher
energy regime the influence of coupling is small, and contrary to
non-tunneling modes, they increase in intensity with increase in energy.
Furthermore, they also do not belong to our previous definition of
tunneling modes. They do not have highest intensity of the same-quanta
group in the anharmonic energy regime and they are not a linear
combination of modes that have a ground state on one of the sites and
occupy high state of the other site. 

\begin{figure}[H]
  \begin{center}
    \includegraphics{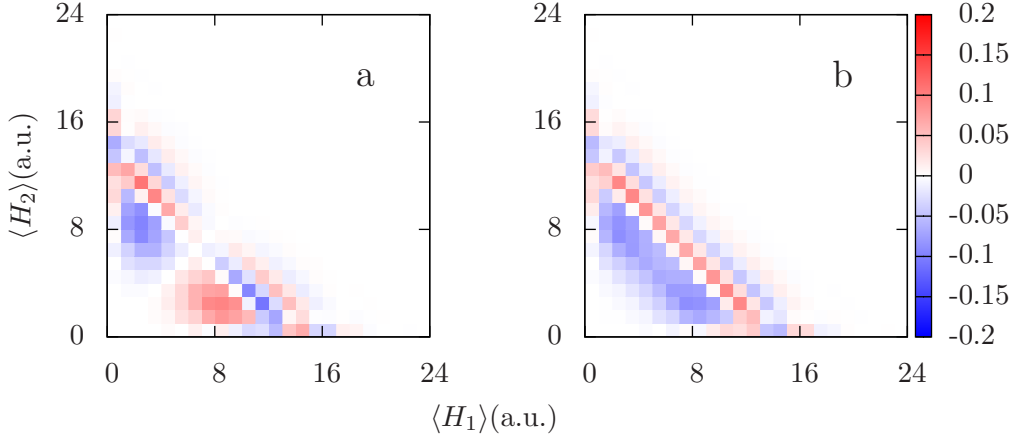}
  \end{center}
  \caption{(Color online) An example of higher order tunneling modes,
    specifically the second order tunneling modes. Contour plots of 90th
    (E=16.01) and 91th (E=16.05) energy levels.}
  \label{FIGhotm}
\end{figure}

The anomaly becomes clear when we analyze at the contour plots of these
eigenstates (Fig.~\ref{FIGhotm}). In these eigenstates, the energy has
to tunnel from one site to the other; it is similar to the tunneling
eigenstates. The contour plots evidence a distinction of these
eigenstates from tunneling modes -- the majority of the wave function
density is not on the axis. These two observations motivate us to define
them as \emph{higher order tunneling modes} (HOTM).  Fig.~\ref{FIGhotm},
specifically, is an example of \emph{second order tunneling modes}
(SOTM).  In the basis set of a direct product of anharmonic oscillators,
they are a linear combination of $|n-1,1\rangle$ and $|1,n-1\rangle$
states.  Fig~\ref{FIGcouunh} shows other HOTM that are even smaller in
intensity than SOTM and manifest at even higher energy limit.  Next, we
compare HOTM and first order tunneling modes and outline a couple of
other interesting points.

To identify the relation of HOTM to tunneling modes, we would like to
point out few similarities. As seen from Fig.~\ref{FIGhotm}, HOTM are a
pair of symmetric and anti-symmetric states with a higher probability of
being on the oscillators rather then shared. On the figures, one can see
the lack of density on the increasing diagonal; the density closer to
axis.  As seen in Fig.~\ref{FIGcou_unharm_crossing__scale_energy} at
the energy 10.33 a.u., HOTM start as degenerate states but further
along, they split into a pair, similarly to the first order tunneling
states. Since we are looking at lower value of coupling, we can see HOTM
at 10.33, rather than at 16 a.u. 


To emphasize that HOTM should be treated different from the first order
tunneling modes, we would like to point out some key differences.  The
probability density in HOTM shifts towards the other oscillator by one
occupational level. Fig.~\ref{FIGhotm} shows this shift in density for
the second order tunneling modes; they correspond to $|n-1,1\rangle$ and
$|1,n-1\rangle$ occupational states. The maximums of the wave-functions
are not on the axes anymore but are closer to the increasing diagonal.
Comparing to the energy region 10.3-10.6 a.u. of the tunneling modes in
Fig.~\ref{FIGcou_unharm_crossing__scale_energy}, the SOTM have larger
splitting between symmetric and anti-symmetric modes than the first
order ones. The distance between the top left and bottom right density
concentrations in the contour plot is related to the tunneling
probability; therefore, it is rational to assume that with decrease of
this distance the splitting and the tunneling probability increases.

\subsubsection{Influence of the Initial Displacement}

Initial state determines the intensities in the spectrum. We show that
when initially both of the sites are displaced, HOTM have higher intensity than
tunneling or non-tunneling modes. The impact of this observation is
twofold: we gain insight on experiments with displaced coherent states
on all sites and we understand influence of HOTM in the dynamics.
Depending on the experimental system setup, the initial conditions can
be different.  When experimental setup cannot target a specific site,
multiple sites will be displaced. An example of this would be
exciting a number of sites in a system from a ground state to an excited
state, where the excited state is displaced. One application may be
understanding of DNA breaking under THz
radiation~\cite{PhysRevE.83.040901}. Emphasizing presence of the HOTM in
the dynamics, on the other hand, results in a new type of dynamics --
the energy is neither localized on sites nor distributed between the
sites.

\begin{figure}[H]
  \begin{center}
    \includegraphics[width=\textwidth]{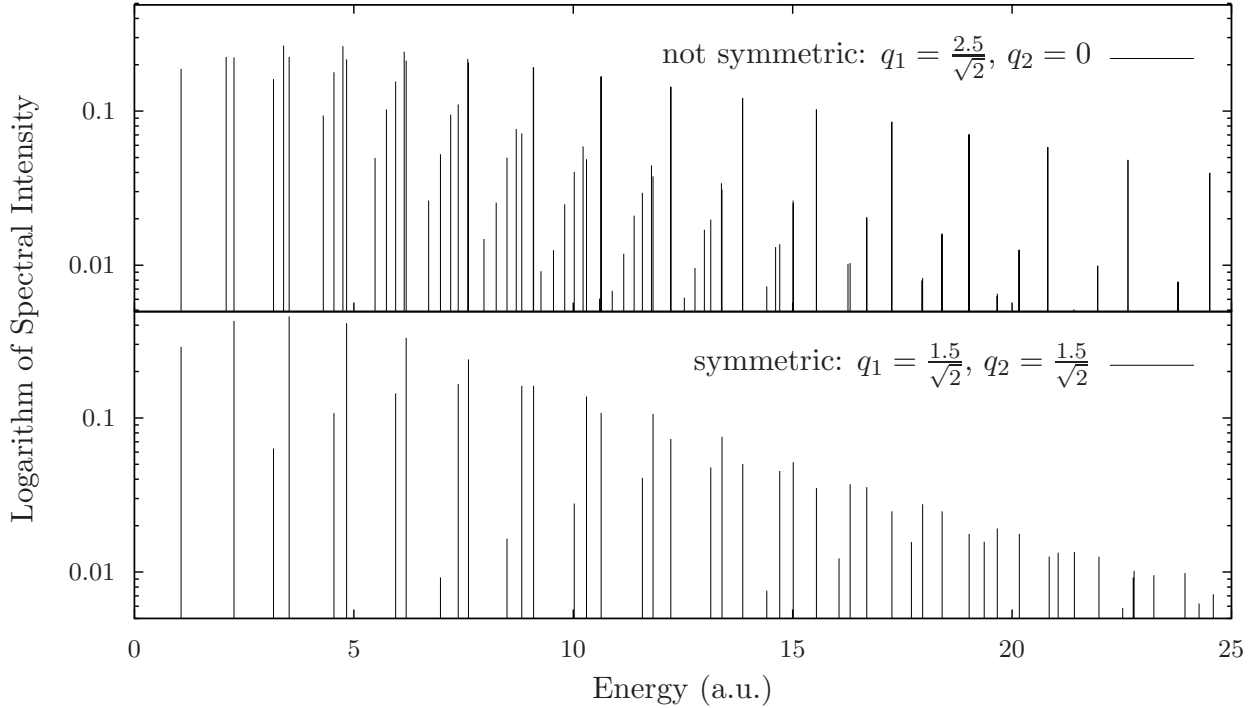}
  \end{center}
  \caption{The \textbf{top} plot is a spectrum from
    Fig.~\ref{FIGautocorrelation} with a log scale. The \textbf{bottom}
    plot is the same as the top plot but for different initial
    conditions -- both of the sites have symmetrically displaced
    coherent state.
  }
  \label{FIGinitial}
\end{figure}

Fig.~\ref{FIGinitial} illustrates the effects of displacing the coherent
states symmetrically on both sites (the bottom panel), in contrast to
displacing on just one of the sites (the top panel).  In the lower
energy region ($<6$ a.u.), the effect of displacing both sites is clear.
At 2 a.u. for example, the case with both sites displaced (BSD) has only
one line corresponding to a symmetric eigenstate; the case with one site
displaced (OSD) has both symmetric and anti-symmetric modes.  In case of
BSD, the pattern continues at higher energies -- only the symmetric
eigenstates have intensity.  Furthermore, in the same-quanta group the
highest energy eigenstate have the highest intensity.  In the transient
region (10-15 a.u.), SOTM start to get priority. The OSD case has an
opposite trend.  At lower energies, middle of the same-quanta group has
the highest intensity. In the transient region, tunneling modes start to
dominate.  However, the most interesting contrast is in the anharmonic
region, where in OSD case, tunneling modes dominate. In BSD, SOTM and
other HOTM dominate.

In classical mechanics if both sites have the same amount of the initial
energy, there is no transfer of energy.  In quantum mechanics on the
other hand, coherent state is a superposition of eigenstates -- the
dynamics is more complex.  To get a brief understanding, let us analyze
the dynamics around 20 a.u.  In the case of OSD, the tunneling mode pair
has most of the intensity.  As we mentioned earlier, the energy
transfered from one site to the other.  In the case of BSD, the
intensity is distributed only among the symmetric modes -- there is no
transfer of energy between the sites.  However, the energy is
distributed almost evenly among HOTM.  HOTM have some density shared
between the sites.  With time, the density oscillates from being
localized on sites to being partially shared between the sites. 

\subsubsection{Autocorrelation of the displaced coherent state}

\begin{figure}[H]
  \begin{center}
    \includegraphics{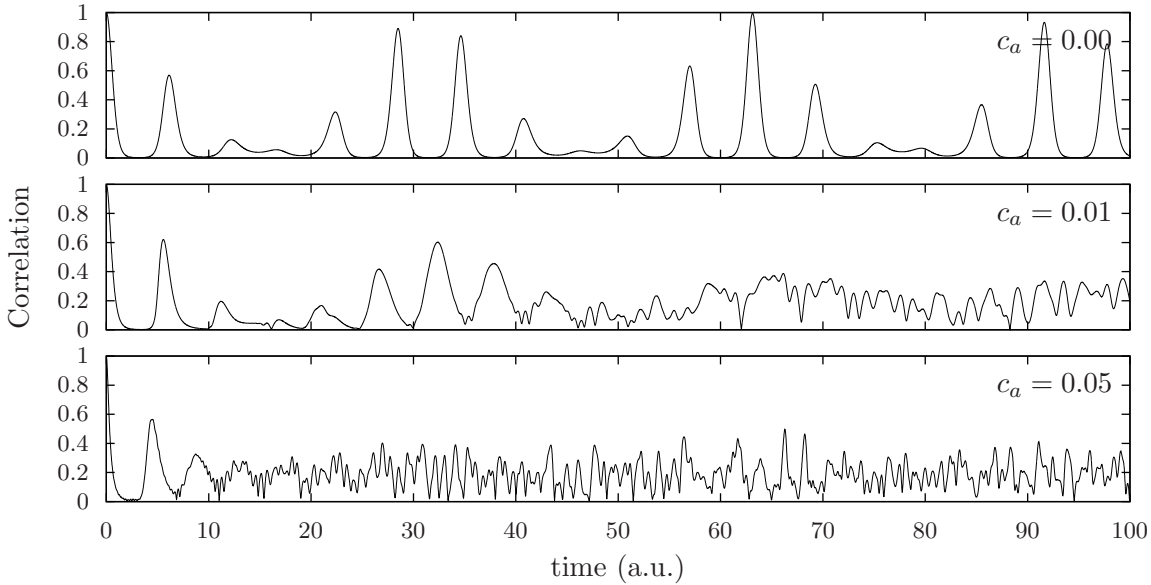}
  \end{center}
  \caption{ Auto-correlation functions $\langle \phi(t)| \phi(0)
  \rangle$ for different value of anharmonicity coefficient $c_a$:
  $c_a=0.00$ (top), $c_a=0.01$ (center) and $c_a=0.05$ (bottom). The
  effects of increasing the anharmonicity and loss of resonance can
  be seen as $c_a$ increases. In all cases the wave packet on the first
  oscillator is displaced to $x_0=1.767$.  }
  \label{FIGautocorrelation}
\end{figure}

A number of time-domain methods are based on propagation of a coherent
states. Examples include HK~\cite{Herman198427},
QHD~\cite{prezhdo:4450\comment{,pereverzev:134107,heatwole:244111,prezhdo:2995,springerlink:10.1007/s00214-005-0032-x,doi:10.1021/jp0709050,Miyachi2006585,Horsfield200816,Ando2003532,heatwole:244111,pahl:8704,prezhdo:2995}},
CSPI~\cite{PhysRevE.69.066204},
CCS~\cite{\comment{halashilin:5367,}shalashilin:244111} and
MP-SOFT~\cite{wu:1676,wu:6720}.  Earlier we calculated eigenstates to
analyze dynamic properties of the system. Eigenstates represent
stationary properties of dynamics. To compare our results to time-domain
methods, we calculate auto-correlation function.  Besides showing
evolution of the initial state in time, auto-correlation function is
used to calculate spectroscopically properties, diffusion rate, and
reaction rate constants.  Fig.~\ref{FIGautocorrelation} shows
auto-correlation function (Eq.\ref{eq:corr}) for different values of
anharmonicity in the potential Eq.(\ref{eq:hamSecondQuant}). The system
is the same as in Fig.~\ref{FIGcouunh} with $c_c=0.2$.

The top panel shows a dynamics of coupled harmonic oscillators. The
wave-function comes back to the previous state with a period of 65 a.u.
Once there is anharmonicity ($c_c=0.01$, middle panel), the eigenstates
have frequencies that are not multiple of each other anymore. The
condition for resonance is lost, and the tails of coherent state move at
different frequencies. The coherent state spreads with time. The wave
function does not come back to the same state. The case with $c_a=0.05$
(bottom panel) is an even more extreme example. The coherent state
spread even faster. Calculating rapidly spreading coherent state is
difficult. This result shows the challenge that coherent state based
path-integral methods have to face -- they have to track quickly
spreading tails of the coherent state. Nevertheless, the time-window
that we used proves to be sufficiently large to get satisfying results.
In addition, quick decrease of auto-correlation function in time
justifies using shorter propagation times.  Time-domain methods that are
based on propagation of coherent state will prove to be an effective and
accurate tool to model signatures of QDBs in multidimensional systems.

\section{Discussion and Conclusions}

Understanding signatures of QDBs in the quantum dynamics of coherent
states is a step towards applying the localization phenomenon to large
scale systems and bridging the gap between concepts of quantum and
classical DBs.  We identified and analyzed tunneling modes, a quantum
counterpart of classical DBs, in the spectrum of quartic dimer using
contour plots. Further, we visualized the eigenstates within transient
region. In contrast to the behavior of classical breathers, a transition
from non-tunneling modes to tunneling modes is gradual and smooth. We
also presented additional important observations: tunneling modes avoid
crossing with non-tunneling modes from the same-quanta group; new types
of modes, HOTM appears at higher energies; controlling initial
conditions allow us to prioritizes HOTM in the spectrum;
auto-correlation function decay faster with increase in non-linearity,
which shortens a required time for calculation.


Tying our results to classical mechanical interpretation of DBs, we show
that tunneling and non-tunneling modes are the quantum mechanical
counter part of localized and delocalized modes that appear in classical
mechanics. Since coherent states span a large number of eigenstates,
they simultaneously contain both tunneling and non-tunneling modes.  In
the classical mechanic analogue, the initial energy localized on one of
the sites, cannot completely transfer to the other site.  In quantum
mechanics, the energy tunnels to the other site; however, as the size of
the system approaches a classical limit, the tunneling time approaches
infinity.

From the analysis of the transient region, we concluded that tunneling
and non-tunneling modes alternate with increasing energy, and that the
transition from non-tunneling to tunneling modes is smooth.  Both of
these conclusions are in contrast with the classical mechanical model of
breathers, where localized and non-local modes are separated sharply at
a certain energy.  The convergence of the transient region in quantum
mechanics to the separatrix point in classical mechanics can be possibly
reached by scaling the quantum result to the classical limit.


Many semi-classical methods that are capable of approaching large
systems rely on propagation of coherent state. We show that both
tunneling and non-tunneling modes are present in the coherent state.
Therefore part of the coherent state would stay localized and part would
be delocalized.  The CSPI methods would have difficulty tracing the
width or other metrics of the coherent state, especially in the
transient region, where both types of states have a close to equal
contribution. Nevertheless, CSPI methods should be able to identity the
location of the transient region.




\section*{Acknowledgments}

The research was funded by the NSF grant CHE-1050405 and NSF CAREER
Award No. 0645340.

\bibliography{bqd}

\end{document}